\documentclass{article}
\usepackage{spconf, amsmath, graphicx, amssymb}

\usepackage{url}

\usepackage{caption}
\usepackage{subcaption}

\title{AUTOENCODER BASED IMAGE COMPRESSION: CAN THE LEARNING BE QUANTIZATION INDEPENDENT?}
\name{Thierry Dumas, Aline, Roumy and Christine Guillemot \thanks{This work has been supported by the French Defense Procurement Agency (DGA).}}
\address{INRIA Rennes Bretagne-Atlantique \\ thierry.dumas@inria.fr, aline.roumy@inria.fr, christine.guillemot@inria.fr}

\begin{document} \sloppy
\maketitle

\begin{abstract}
This paper explores the problem of learning transforms for image compression via autoencoders. Usually, the rate-distortion performances of image compression are tuned by varying the quantization step size. In the case of autoencoders, this in principle would require learning one transform per rate-distortion point at a given quantization step size. Here, we show that comparable performances can be obtained with a unique learned transform. The different rate-distortion points are then reached by varying the quantization step size at test time. This approach saves a lot of training time.
\end{abstract}


\section{Introduction} \label{sec:1}
Image coding standards all use linear and invertible transforms to convert an image into coefficients with low statistical dependencies, i.e suited for scalar quantization. Notably, the discrete cosine transform (DCT) is the most commonly used for two reasons: $(i)$ it is image-independent, implying that the DCT does not need to be transmitted, $(ii)$ it approaches the optimal orthogonal transform in terms of rate-distortion, assuming that natural images can be modeled by zero-mean Gaussian-Markov processes with high correlation \cite{source_coding_part}. Deep autoencoders have been shown as promising tools for finding alternative transforms \cite{variable_rate_image, towards_conceptual_compression, image_compression_with}. Autoencoders learn the encoder-decoder non-linear transform from natural images.

In the best image compression algorithms based on autoencoders \cite{lossy_image_compression, end_to_end, real_time_adaptive}, one transform is learned per rate-distortion point at a given quantization step size. Then, the quantization step size remains unchanged at test time so that the training and test conditions are identical. By contrast, image coding standards implement adaptive quantizations \cite{an_overview_of, video_coding_part}. Should the quantization be imposed during the training? To answer this, we propose an approach where the transform and the quantization are learned jointly. Then, we investigate whether, at test time, the compression falls apart when the coefficients obtained with the learned transform are quantized using quantization step sizes which differ from those in the training stage. The code to reproduce our numerical results and train the autoencoders is available online\footnote{\url{www.irisa.fr/temics/demos/visualization_ae/visualizationAE.htm} \label{foot}}.

Matrices and tensors are denoted by bold letters. $\left\Vert \mathbf{X} \right\Vert_{F}$ is the Frobenius norm of $\mathbf{X}$. $\mathbf{X} \odot \mathbf{Z}$ is the elementwise multiplication between $\mathbf{X}$ and $\mathbf{Z}$.

\section{Joint learning of the transform and the quantization} \label{sec:2}
Section \ref{sec:2} introduces an efficient autoencoder for image compression. Then, it details our proposal for learning jointly this autoencoder transform and the quantization.

\subsection{Autoencoder for image compression} \label{subsec:2.1}
\begin{figure*}
	\centering
	\includegraphics[trim=0.8cm 11.2cm 0.8cm 0cm, clip, width=0.80\linewidth]{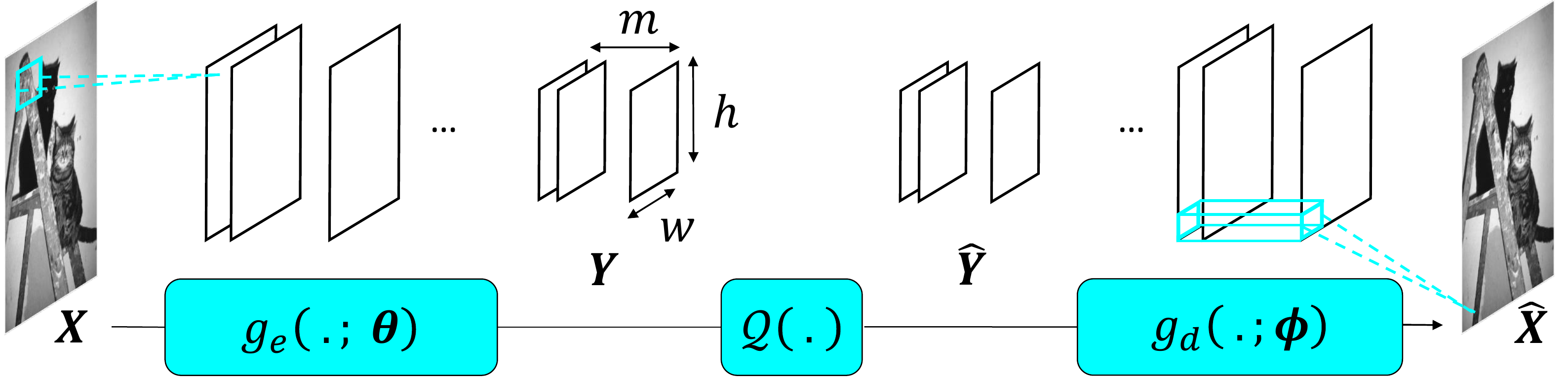}
	\vspace{-0.6 \baselineskip}
	\caption{Illustration of an autoencoder for image compression.}
	\vspace{-0.7 \baselineskip}
	\label{fig:1}
\end{figure*}

An autoencoder is a neural network with an encoder $g_{e}$, parametrized by $\boldsymbol{\theta}$, that computes a representation $\mathbf{Y}$ from the data $\mathbf{X}$, and a decoder $g_{d}$, parametrized by $\boldsymbol{\phi}$, that gives a reconstruction $\hat{\mathbf{X}}$ of $\mathbf{X}$, see Figure \ref{fig:1}. Autoencoders can be used for denoising or dimensionality reduction. When it is used for compression, the representation is also quantized, leading to the new quantized representation $\hat{\mathbf{Y}} = \mathcal{Q} \left( \mathbf{Y} \right)$. If an autoencoder has fully-connected layers \cite{semantic_hashing, binary_coding_of, using_very_deep}, the number of parameters depends on the image size. This implies that one autoencoder has to be trained per image size. To avoid this, an architecture without fully-connected layer is chosen. It exclusively comprises convolutional layers and non-linear operators. In this case, $\mathbf{Y} \in \mathbb{R}^{h \times w \times m}$ is a set of $m$ feature maps of size $n = h \times w$, see Figure \ref{fig:1}.

The basic autoencoder training minimizes the image reconstruction error \cite{learning_deep_architectures}. In order to create a rate-distortion optimization, the authors in \cite{end_to_end} add the minimization of the entropy of the quantized representation. Moreover, a bit allocation is performed by learning a normalization for each feature map of $\mathbf{Y}$. The encoder followed by the normalizations at the encoder side, parametrized by $\boldsymbol{\varphi}_{e}$, are denoted $\overline{g}_{e} \left( \; . \; ; \boldsymbol{\theta}, \boldsymbol{\varphi}_{e} \right)$. Similarly, the normalizations at the decoder side, parametrized by $\boldsymbol{\varphi}_{d}$, followed by the decoder are denoted $\overline{g}_{d} \left( \; . \; ; \boldsymbol{\varphi}_{d}, \boldsymbol{\phi} \right)$. Finally, this leads to \eqref{eq:1}.

\begin{align} \label{eq:1}
	\min\limits_{\substack{\boldsymbol{\theta}, \boldsymbol{\varphi}_{e}, \\ \boldsymbol{\varphi}_{d}, \boldsymbol{\phi}}} \: \mathbb{E} &\left[ \left\Vert \mathbf{X} - \overline{g}_{d} \left( \mathcal{Q} \left( \overline{g}_{e} \left( \mathbf{X}; \boldsymbol{\theta}, \boldsymbol{\varphi}_{e} \right) \right); \boldsymbol{\varphi}_{d}, \boldsymbol{\phi} \right) \right\Vert_{F}^{2} + \gamma \sum \limits_{i = 1}^{m} H_{i} \right] \nonumber \\
	H_{i} &= - \frac{1}{n} \sum \limits_{j = 1}^{n} \log_{2} \left( \hat{p}_{i} \left( \hat{y}_{ij} \right) \right), \gamma \in \mathbb{R}_{+}^{*}
\end{align}
$\hat{p}_{i}$ is the probability mass function of the $i^{\text{th}}$ quantized feature map coefficients $\{ \hat{y}_{ij} \}_{j = 1 ... n}$. The expectation $\mathbb{E} [ . ]$ is approximated by averaging over a training set of images. Unfortunately, $\mathcal{Q}$ makes minimization \eqref{eq:1} unusable. Indeed, the derivative of any quantization with respect to its input is $0$ at any point. Consequently, $\boldsymbol{\theta}$ and $\boldsymbol{\varphi}_{e}$ cannot be learned via gradient-based methods \cite{learning_representations_by}. To get around this issue, \cite{end_to_end} fixes the quantization step size to 1 and approximates the uniform scalar quantization with the addition of a uniform noise of support [-0.5, 0.5]. Note that, even though the quantization step size is fixed, the bit allocation varies over the different feature maps via the normalizations. In the next section, we consider instead to remove the normalizations and learn explicitly the quantization step size for each feature map of $\mathbf{Y}$.

\subsection{Learning the quantization step sizes} \label{subsec:2.2}
We address the problem of optimizing the quantization step size for each feature map of $\mathbf{Y}$. Because of the quantization, the function to be minimized is an implicit function of the quantization step sizes $\{ \delta_{i} \}_{i = 1 ... m}$. The target is to make it an explicit function of $\{ \delta_{i} \}_{i = 1 ... m}$. For $q \in \{ ..., -\delta_{i}, 0, \delta_{i}, ... \}$,
\begin{align} \label{eq:2}
	\hat{p}_{i} \left( q \right) = \int_{q - 0.5\delta_{i}}^{q + 0.5\delta_{i}} p_{i} \left( t \right) dt = \delta_{i} \tilde{p}_{i} \left( q \right)
\end{align}
$\tilde{p}_{i} = p_{i} \ast l_{i}$ where $p_{i}$ is the probability density function of the $i^{\text{th}}$ feature map coefficients $\{ y_{ij} \}_{j = 1 ... n}$ and $l_{i}$ denotes the probability density function of the continuous uniform distribution of support $[-0.5 \delta_{i}, 0.5 \delta_{i}]$. The normalizations are removed from \eqref{eq:1} and, using \eqref{eq:2}, \eqref{eq:1} becomes \eqref{eq:3}.
\begin{align} \label{eq:3}
	\min\limits_{\boldsymbol{\theta}, \boldsymbol{\phi}} \: \mathbb{E} \Bigg[ &\left\Vert \mathbf{X} - g_{d} \left( g_{e} \left( \mathbf{X}; \boldsymbol{\theta} \right) + \boldsymbol{\mathcal{E}}; \boldsymbol{\phi} \right) \right\Vert_{F}^{2} + \gamma \sum \limits_{i = 1}^{m} \tilde{h}_{i} \Bigg]\\
	\tilde{h}_{i} = &- \log_{2} \left( \delta_{i} \right) - \frac{1}{n} \sum \limits_{j = 1}^{n} \log_{2} \left( \tilde{p}_{i} \left( y_{ij} + \varepsilon_{ij} \right) \right) \nonumber
\end{align}
The $i^{\text{th}}$ matrix of $\boldsymbol{\mathcal{E}} \in \mathbb{R}^{h \times w \times m}$ contains $n$ realizations $\{ \varepsilon_{ij} \}_{j = 1 ... n}$ of $\mathcal{E}_{i}$, $\mathcal{E}_{i}$ being a continuous random variable of probability density function $l_{i}$. In \eqref{eq:3}, the function to be minimized is differentiable with respect to $\boldsymbol{\theta}$. $\boldsymbol{\theta}$ can thus be learned via gradient-based methods. However, $\{ \delta_{i} \}_{i = 1 ... m}$ cannot yet be learned as the function to be minimized in \eqref{eq:3} is not differentiable with respect to $\{ \delta_{i} \}_{i = 1 ... m}$. This is resolved using the change of variable $\mathcal{E}_{i} = \delta_{i} \mathcal{T}$ where $\mathcal{T}$ is a random variable following the continuous uniform distribution of support $[-0.5, 0.5]$. Now, the minimization over $\{ \delta_{i} \}_{i = 1 ... m}$ is feasible, see $\eqref{eq:4}$.
\begin{align} \label{eq:4}
	\min\limits_{\substack{\boldsymbol{\theta}, \boldsymbol{\phi}, \\ \delta_{1}, ..., \delta_{m}}} \: \mathbb{E} \Bigg[ &\left\Vert \mathbf{X} - g_{d} \left( g_{e} \left( \mathbf{X}; \boldsymbol{\theta} \right) + \mathbf{\Delta} \odot \mathbf{T}; \boldsymbol{\phi} \right) \right\Vert_{F}^{2} + \gamma \sum \limits_{i = 1}^{m} \tilde{h}_{i} \Bigg] \nonumber\\
	\tilde{h}_{i} = &- \log_{2} \left( \delta_{i} \right) - \frac{1}{n} \sum \limits_{j = 1}^{n} \log_{2} \left( \tilde{p}_{i} \left( y_{ij} + \delta_{i} \tau_{ij} \right) \right)
\end{align}
The $i^{\text{th}}$ matrix of $\mathbf{T} \in \mathbb{R}^{h \times w \times m}$ contains $n$ realizations $\{ \tau_{ij} \}_{j = 1 ... n}$ of $\mathcal{T}$. All the coefficients in the $i^{\text{th}}$ matrix of $\boldsymbol{\Delta} \in \mathbb{R}^{h \times w \times m}$ are equal to $\delta_{i}$. A detail has been left out so far: $\tilde{p}_{i}$ is unknown. In a similar manner to \cite{lossy_image_compression, end_to_end}, $\tilde{p}_{i}$ can be replaced by a function $\tilde{f}_{i}$, parametrized by $\boldsymbol{\psi}^{(i)}$, and $\boldsymbol{\psi}^{(i)}$ is learned such that $\tilde{f}_{i}$ fits $\tilde{p}_{i}$.

In the end, we end up with three groups of parameters: $\{ \boldsymbol{\theta}$, $\boldsymbol{\phi} \}$, $\{ \delta_{i} \}_{i = 1 ... m}$ and $ \{ \boldsymbol{\psi}^{(i)} \}_{i = 1 ... m}$. These three groups are learned by alternating three different stochastic gradient descents. All the training heuristics are detailed in the code\textsuperscript{\ref{foot}}.

Section \ref{sec:2} has developped an approach for learning explictly the transform and a quantization step size for each feature map of $\mathbf{Y}$. Before evaluating this approach in Section \ref{sec:4}, Section \ref{sec:3} studies what would happen if, at test time, the coefficients in $\mathbf{Y}$ are quantized using quantization step sizes that differ from those in the training stage. This first requires understanding the internal structure of $\mathbf{Y}$ after the training.

\section{Inside the learned representation} \label{sec:3}
This section studies the different feature maps of $\mathbf{Y}$ after the training. To this end, a deep convolutional autoencoder must first be built and trained. $g_{e}$ is the composition of a convolutional layer, a generalized divisive normalization (GDN) \cite{density_modeling_of}, a convolutional layer, a GDN and a convolutional layer. $g_{d}$ is the reverse composition, replacing each GDN with an inverse generalized divisive normalization (IGDN) \cite{density_modeling_of} and each convolutional layer with a transpose convolutional layer \cite{a_guide_to}. It is important to stress that $m = 128$, $\mathbf{X}$ has one channel and the convolutional strides and paddings are chosen such that $h$ and $w$ are 16 times smaller than respectively the height and the width of $\mathbf{X}$. Therefore, the number of pixels in $\mathbf{X}$ is twice the number of coefficients in $\mathbf{Y}$. The training set contains $24000$ luminance images of size $256 \times 256$ that are extracted from ImageNet \cite{imagenet_a_large}. The minimization is \eqref{eq:4}, $\gamma = 10000.0$. Note that,  if a GDN was placed immediately after $g_{e}$, a IGDN was placed immediately before $g_{d}$ and, $\forall i \in [|1, m|], \delta_{i} = 1.0$ was not learned, the autoencoder architecture and the training would correspond to \cite{end_to_end}.

\subsection{Distribution of the learned representation} \label{subsec:3.1}
\begin{figure}
	\centering
	\begin{subfigure}{0.23\textwidth}
		\includegraphics[width=\linewidth]{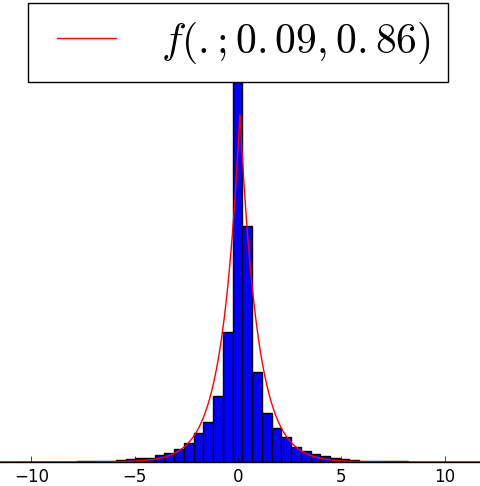}
		\caption{$i = 50$}
	\end{subfigure}
	\begin{subfigure}{0.23\textwidth}
		\includegraphics[width=\linewidth]{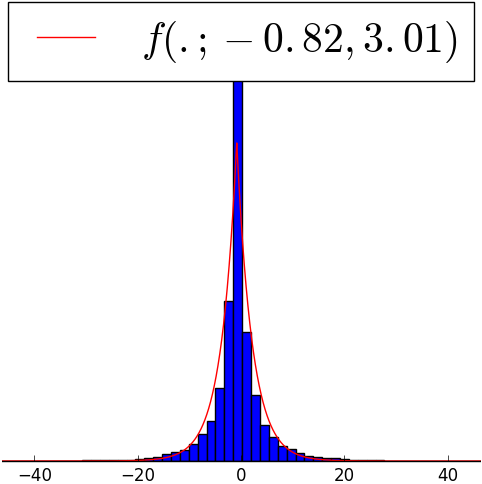}
		\caption{$i = 125$}
	\end{subfigure}
	\vspace{-0.4 \baselineskip}
	\caption{Normed histogram of the $i^{\text{th}}$ feature map of $\mathbf{Y}$.}
	\vspace{-0.5 \baselineskip}
	\label{fig:2}
\end{figure}
\begin{figure}
	\centering
	\includegraphics[width=0.35\textwidth]{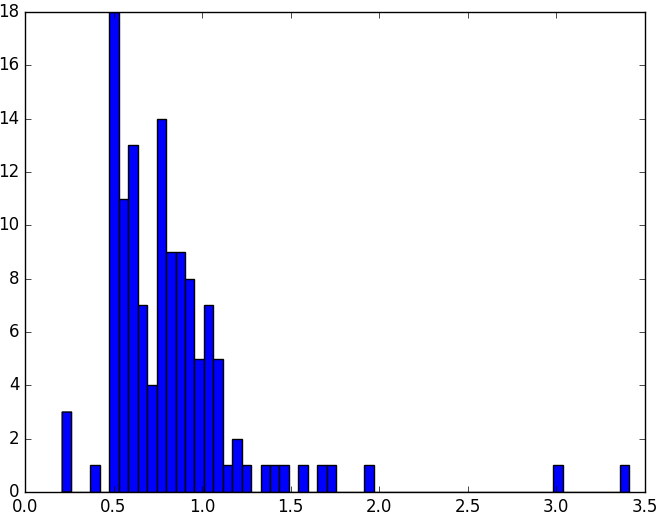}
	\vspace{-0.5 \baselineskip}
	\caption{Histogram of the $m - 1$ scales provided by the fitting.}
	\vspace{-0.5 \baselineskip}
	\label{fig:3}
\end{figure}
After the training, a test set of $24$ luminance images of size $512 \times 768$ is created from the Kodak suite\footnote{\textit{r0k.us/graphics/kodak/}}. Here, $\mathbf{X}$ refers to a test luminance image. Figure \ref{fig:2} shows the normed histogram of the $50^{\text{th}}$ feature map of $\mathbf{Y} = g_{e} \left( \mathbf{X}; \boldsymbol{\theta} \right)$ and that of its $125^{\text{th}}$ feature map, averaged over the test set. Every feature map of $\mathbf{Y}$, except the $90^{\text{th}}$, has a normed histogram similar to those displayed. To be more precise, let's write the probability density function of the Laplace distribution with mean $\mu \in \mathbb{R}$ and scale $\lambda \in \mathbb{R}_{+}^{*}$, denoted $f \left( \; . \; ; \mu, \lambda \right)$.

\begin{align*}
	f \left( x; \mu, \lambda \right) = \frac{1}{2\lambda} \exp \left(- \frac{\left\vert x - \mu \right\vert}{\lambda} \right)
\end{align*}
$\forall i \in [|1, m|]$, $i \neq 90$, there exists $\mu_{i} \in \mathbb{R}$ and $\lambda_{i} \in \mathbb{R}^{*}_{+}$ such that $f \left( \; . \; ; \mu_{i}, \lambda_{i} \right)$ fits well the normed histogram of the $i^{\text{th}}$ feature map of $\mathbf{Y}$. Note that most of the $m - 1$ scales belong to $[0.5, 2.0]$, see Figure \ref{fig:3}. For transformed coefficients having a zero-mean Laplace distribution, \cite{efficient_scalar_quantization} proves that a uniform reconstruction quantizer (URQ) with constant decision offsets approaches the optimal scalar quantizer in terms of squared-error distortion for any quantization step size. Yet, in our case, $(i)$ the $m - 1$ Laplace probability density functions are not zero-mean, $(ii)$ uniform scalar quantizers are used instead of this URQ. The point $(i)$ is not problematic as an extra set of luminance images is used to compute an approximation $\overline{\mu}_{i} \in \mathbb{R}$ of the mean of the $i^{\text{th}}$ feature map of $\mathbf{Y}$, then, at test time, the $i^{\text{th}}$ feature map of $\mathbf{Y}$ is centered via $\overline{\mu}_{i}$ before being quantized. Note that $\{ \overline{\mu_{i}} \}_{i = 1 ... m}$ does not depend on the test luminance images, thus incurring no transmission cost. Regarding the point $(ii)$, it must be noted that the decoder mapping of the URQ is exactly the decoder mapping of the uniform scalar quantization with same quantization step size. Since our case comes close to the requirements of the proof in \cite{efficient_scalar_quantization}, at test time, the rate-distortion trade-off should not collapse as the quantization step sizes deviate from the learned values. This will be verified in Section \ref{sec:4}.

\subsection{Internal structure of the learned representation} \label{subsec:3.2}
\begin{figure}
	\centering
	\begin{subfigure}[t]{0.11\textwidth}
		\includegraphics[width=\textwidth]{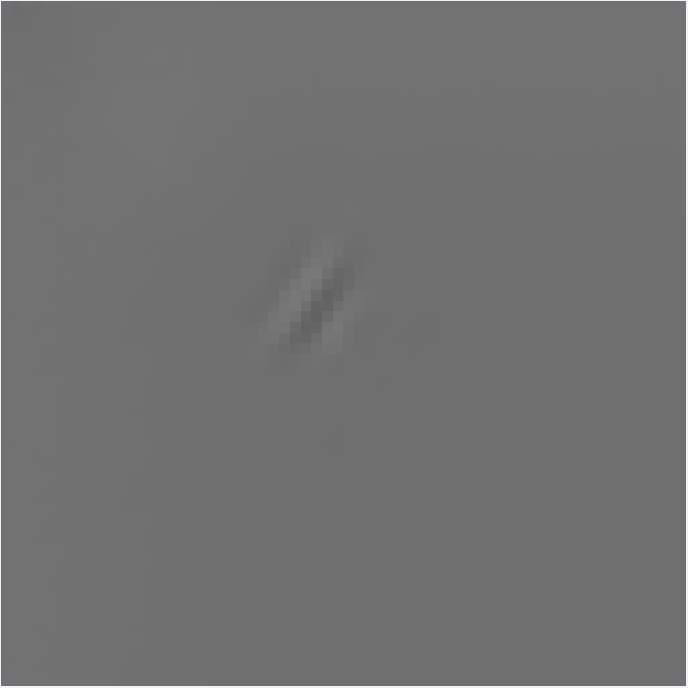}
		\caption{$\alpha = 8.0$}
	\end{subfigure}
	\begin{subfigure}[t]{0.11\textwidth}
		\includegraphics[width=\textwidth]{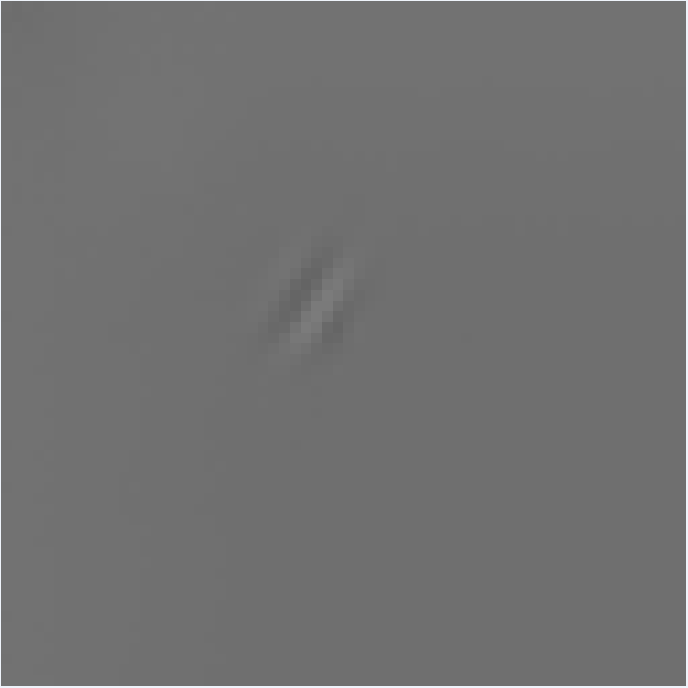}
		\caption{$\alpha = -8.0$}
	\end{subfigure}
	\begin{subfigure}[t]{0.11\textwidth}
		\includegraphics[width=\textwidth]{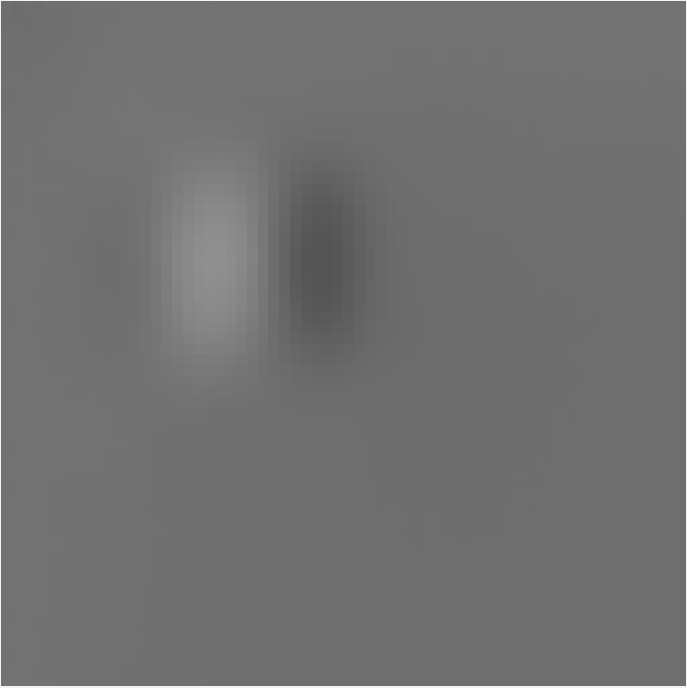}
		\caption{$\alpha = 20.0$}
	\end{subfigure}
	\begin{subfigure}[t]{0.11\textwidth}
		\includegraphics[width=\textwidth]{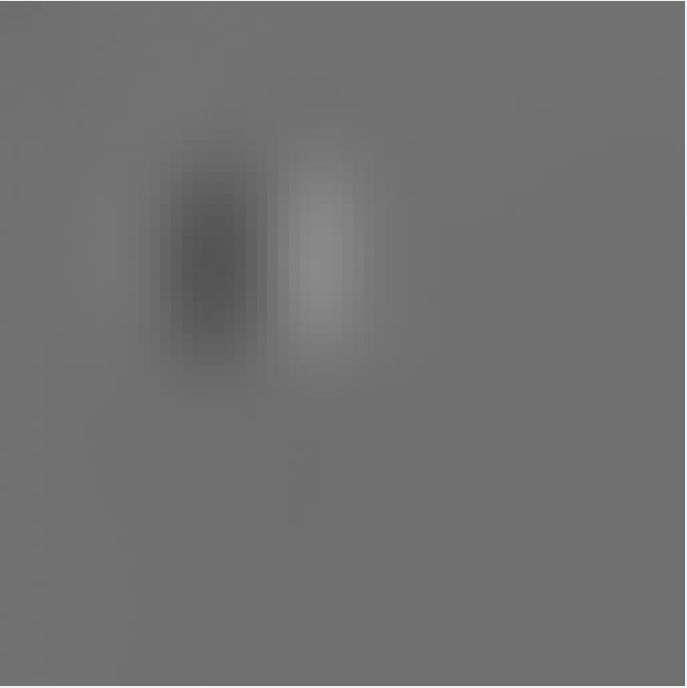}
		\caption{$\alpha = -20.0$}
	\end{subfigure}
	\caption{$64 \times 64$ crop at the top-left of $\hat{\mathbf{X}}$. $j = 50$ in $(a)$ and $(b)$. $j =125$ in $(c)$ and $(d)$.}
	\label{fig:4}
\end{figure}
The shortcoming of the previous fitting is that it does not reveal what information each matrix of $\mathbf{Y}$ encodes. To discover it, further visualizations are needed. The most common way of exploring a deep convolutional neural network (CNN) trained for image recognition is to look at the image, at the CNN input, resulting from the maximization over its pixels of a given neural activation in the CNN \cite{deep_inside_convolutional, deep_neural_networks, understanding_neural_networks}. Precisely, \cite{deep_inside_convolutional, deep_neural_networks} maximize over the image pixels a given neural activation at the CNN output, i.e a class probability. This shows what image features characterize this class according to the CNN. In our case, the maximization over the image pixels of a given coefficient in $\mathbf{Y}$ does not yield interpretable images. Indeed, the coefficients in $\mathbf{Y}$ are not bounded. This may explain why the maximization often returns saturated images.

Alternatively, the information the $j^{\text{th}}$ feature map of $\mathbf{Y}$ encodes, $j \in [|1, m|]$, can be seen as follows. $\forall i \in [|1, m|]$, all the coefficients in the $i^{\text{th}}$ feature map of $\mathbf{Y}$ are set to $\overline{\mu}_{i}$. This way, the feature maps of $\mathbf{Y}$ contains no significant information. Then, a single coefficient in the $j^{\text{th}}$ feature map of $\mathbf{Y}$ is set to $\alpha \in \mathbb{R}$ and $\hat{\mathbf{X}} = g_{d} \left( \mathcal{Q} \left( \mathbf{Y} \right); \boldsymbol{\phi} \right)$ is displayed. $\alpha$ is selected such that it is near one of the two tails of the Laplace distribution of the $j^{\text{th}}$ feature map of $\mathbf{Y}$. Figure \ref{fig:4} shows the $64 \times 64$ crop at the top-left of $\hat{\mathbf{X}}$ when the single coefficient is located at the top-left corner of the $j^{\text{th}}$ feature map of $\mathbf{Y}$, $j \in \{ 50, 125 \}$. We see that the $50^{\text{th}}$ feature map of $\mathbf{Y}$ encodes a spatially localized image feature whereas its $250^{\text{th}}$ feature map encodes a spatially extended image feature. Moreover, the image feature is turned into its symmetrical feature, with respect to the mean pixel intensity, by moving $\alpha$ from the right tail of the Laplace distribution of the $j^{\text{th}}$ feature map of $\mathbf{Y}$ to the left tail. This linear behaviour is observed for each feature map of $\mathbf{Y}$.

It is interesting to see that, given the fitting in Section \ref{subsec:3.1}, $\mathbf{Y}$ is similar to the DCT coefficients for blocks of prediction error samples in H.265 \cite{video_coding_part} in terms of distribution. However, when looking at the information each feature map of $\mathbf{Y}$ encodes, $\mathbf{Y}$ has nothing to do with these DCT coefficients.

\section{Experiments} \label{sec:4}
\begin{figure}
	\centering
	\vspace{-0.2 \baselineskip}
	\includegraphics[width=0.48\textwidth]{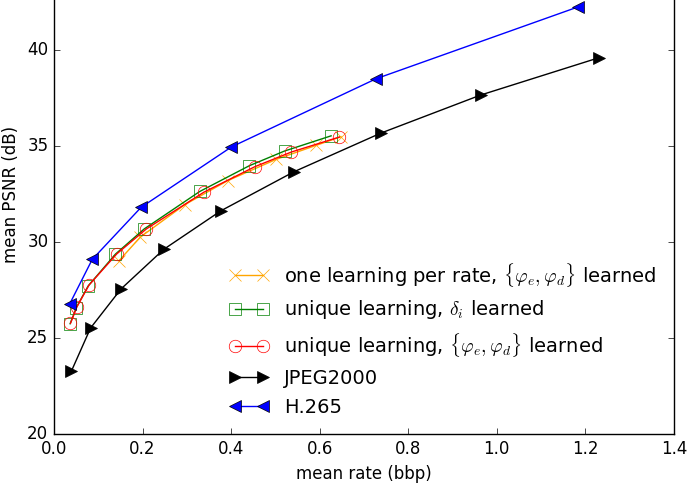}
	\vspace{-1.8 \baselineskip}
	\caption{Rate-distortion curves averaged over the $24$ luminance images from the Kodak suite.}
	\vspace{-1.4 \baselineskip}
	\label{fig:5}
\end{figure}
We now evaluate in terms of rate-distortion performances: $(i)$ whether the way of learning the quantization matters, $(ii)$ whether, at test time, it is efficient to quantize the coefficients obtained with the learned transform using quantization step sizes which differ from those in the training stage. This is done by comparing three cases.

The $1^{\text{st}}$ case follows the approach in \cite{end_to_end}. One transform is learned per rate-distortion point, the bit allocation being learned via the normalizations. In details, an autoencoder is trained for each $\gamma \in S = \{ 10000.0, 12000.0, 16000.0$, $24000.0, 40000.0, 72000.0, 96000.0 \}$. During the training and at test time, the quantization step size is fixed to 1.0.

In the $2^{\text{nd}}$ case, a unique transform is learned, the bit allocation being done by learning a quantization step size per feature map. More precisely, a single autoencoder is trained for $\gamma = 10000.0$ and $\{ \delta_{i} \}_{i = 1 ... m}$ is learned, see Section \ref{sec:2}. At test time, the rate varies as the quantization step sizes are equal to the learned quantization step sizes multiplied by $\beta \in \mathcal{B} = \{ 1.0, 1.25, 1.5, 2.0, 3.0, 4.0, 6.0, 8.0, 10.0 \}$.

In the $3^{\text{rd}}$ case, a unique transform is learned, the bit allocation being learned via the normalizations. In details, a single autoencoder is trained for $\gamma = 10000.0$ and, during the training, the quantization step size is 1.0. At test time, the rate varies as the quantization step size spans $\mathcal{B}$.

In the $2^{\text{nd}}$ case, the autoencoder has the architecture described at the beginning of Section \ref{sec:3}. In the $1^{\text{st}}$ and $3^{\text{rd}}$ case, a GDN is also placed after $g_{e}$ and a IGDN is placed before $g_{d}$. The autoencoders are trained on $24000$ luminance images of size $256 \times 256$ that are extracted from ImageNet. Then, at test time, the $24$ luminance images from the Kodak suite are inserted into the autoencoders. The rate is estimated via the empirical entropy of the quantized coefficients, assuming that the quantized coefficients are i.i.d. Note that, for the $2^{\text{nd}}$ and the $3^{\text{rd}}$ case, we have also implemented a binarizer and a binary arithmetic coder to compress the quantized coefficients losslessly, see the code\textsuperscript{\ref{foot}}. The difference between the estimated rate and the exact rate via the lossless coding is always smaller than $0.04$ bbp. Figure \ref{fig:5} shows the rate-distortion curves averaged over the $24$ luminance images. The JPEG2000 curve is obtained using ImageMagick. The H.265 \cite{overview_of_the} curve is computed via the version HM-16.15. There is hardly any difference between the $2^{\text{nd}}$ and the $3^{\text{rd}}$ case. This means that the explicit learning of the transform and the quantization step sizes is equivalent to learning the transform and the normalizations while the quantization step size is imposed. Note that, in the $2^{\text{nd}}$ case, the learning of $\{ \delta_{i} \}_{i = 1 ... m}$ involves $128$ parameters whereas, in the $3^{\text{rd}}$ case, that of $ \{ \boldsymbol{\varphi}_{e}, \boldsymbol{\varphi}_{d} \}$ involves $33024$ parameters. The $2^{\text{nd}}$ and the $3^{\text{rd}}$ case perform as well as the $1^{\text{st}}$ case. The minimization \eqref{eq:4} and the training in \cite{end_to_end} provide learned transforms which can be used with various quantization step sizes at test time. It is convenient not to train one autoencoder per compression rate as a single training takes 4 days on a NVIDIA GTX 1080. Finally, we see that the learned transforms yield better rate-distortion performances than JPEG2000. The quality of image reconstruction for the experiment in Figure \ref{fig:5} and another experiment on luminance images created from the BSDS300 \cite{a_database_of} can be seen online\textsuperscript{\ref{foot}}.

\section{Conclusion} \label{sec:5}
Using a unique transform learned via autoencoders and various quantization step sizes at test time, it is possible to compress as well as when learning one transform per rate-distortion point at a given quantization step size. Moreover, the learned transformed outperform other image compression algorithms based on transforms.

\bibliographystyle{IEEE}
\bibliography{autoencoder_based}

\end{document}